\documentclass[12pt,preprint]{aastex}

\slugcomment{Submitted to the Astron. J.}

\shorttitle{Simple Analytic Orbits}
\shortauthors{Struck}

\begin{document}


\title{Simple, Accurate, Approximate Orbits in the Logarithmic and \\
    a Range of Power-Law Galactic Potentials}


\author{Curtis Struck}
\affil{Physics and Astronomy Department, Iowa State University,
    Ames, IA 50011}
\email{curt@iastate.edu}



\begin{abstract}

Curves in a family derived from powers of the polar coordinate formula
for ellipses are found to provide good fits to bound orbits in a range
of power-law potentials. This range includes the well-known $1/r$
(Keplerian) and logarithmic potentials. These approximate orbits,
called p-ellipses, retain some of the basic geometric properties of
ellipses. They satisfy and generalize Newton's apsidal precession
formula, which is one of the reasons for their surprising
accuracy. Because of their simplicity the p-ellipses make very useful
tools for studying trends among power-law potentials, and especially
the occurence of closed orbits. The occurence of closed or nearly
closed orbits in different potentials highlights the possibility of
period resonances between precessing, eccentric orbits and circular orbits,
or between the precession period of multi-lobed closed orbits and
satellite periods. These orbits and their resonances promise to help
illuminate a number of problems in galaxy and accretion disk dynamics.

\end{abstract}


\keywords{stellar dynamics ---
galaxies: halos ---
galaxies: kinematics and dynamics}


\section{Introduction}

In recent decades, the recognition that galaxies are surrounded by
extended halos of dark matter has spurred a greatly increased interest
in the structure of orbits in gravitational potentials that are
shallower than the Keplerian $1/r$ potential (e.g., \citealt{bin87},
\citealt{ber00}). Similarly, the mass distribution in galactic bulges
and elliptical galaxies is also extended, and the potential shallower
than the Keplerian one. The logarithmic potential, $\Phi \propto
\ln(r)$, is among the simplest and most studied examples of such a
shallow potential, and will be used as a primary example in this
work. The celestial mechanics of such potentials is not nearly as well
developed as the centuries old literature of the classical
Kepler-Newton potential, but there has been much recent progress (see
\citealt{boc96} and \citealt{con02}).

The orbits in most physically relevant potentials can be numerically
integrated very quickly and accurately, except near singularities and
resonances. Direct numerical integration has provided much information
about orbit families in both general symmetric, and non-axisymmetric
potentials relevant to barred galaxies (including non-axisymmetric
logarithmic potentials, see e.g., \citealt{mir89}).

There has also been much progress in the areas of analytic
approximation and global properties of orbits in non-Keplerian
potentials, and power-law potentials in particular. Concerning global
properties, \citet{sto03} have studied the topology of energy surfaces
of the logarithmic potential and derived a complete qualitative
description of orbits in the axisymmetric case. They also proved that
orbits are centrophobic for the non-axisymmetric case as
well. \citet{val05} have presented extensions of Newton's apsidal
precession theorem \citep{new87} for power-law potentials, and in the
process found no evidence for cases of zero precession at high
eccentricities. The work below provides further support for that
conjecture.

With regard to analytic approximation, we note first of all that
classical epicyclic approximations are the most well developed (e.g.,
see \citealt{bin87}, \citealt{deh99}, and \citealt{con02}). Beyond
this, \citet{con90} have used the multiple Fourier series method of
\citet{pre82} to find orbit families and to derive accurate orbit
approximations for the symmetric and non-axisymmetric logarithmic
potentials. \citet{ada05} have examined spirograph or epicycloid orbit
approximations and found that they accurately fit a wide range of
orbits in the Hernquist and other halo potentials (with only two
``spirograph'' circles). The spirograph or epicycloid approximation is
formally distinct from the epicyclic approximation, though similar in
that they both can be extended to arbitrary accuracy.

More recently, \citet{tou97} developed a relatively simple
symplectic mapping for non-axisymmetric potentials that captures the
behavior of orbits in these potentials. The map is based on the
approximation that orbital evolution is driven by two processes:
apsidal precession in the axisymmetric part of the potential, and
torques by the non-axisymmetric potential, concentrated at
apoapse. Because this symplectic map is based on apsidal precession,
the Valluri et al. paper and the results below on precession
complement it.

With this recent work the knowledge base on orbits in galaxy
potentials (and similarly shallow potentials in galaxy groups and
clusters) has become quite extensive. However, the basis of much of
classical celestial mechanics (and many specific approximation
schemes) is the simplicity of Kepler's elliptical orbit solution. The
conceptual and analytical foundation of galactic dynamics and other
fields that use power-law potentials is inevitably weaker without such
simple orbital solutions. Unfortunately, mechanics textbooks teach us
that analytic orbital solutions only exist in a few special cases
among the power-law forces (\citealt{gol80}, \citealt{dan88}). The
simple epicyclic approximation provides a versatile computational
tool, but with limitations as a conceptual tool if multiple or large
epicycles are needed for accurate calculations (e.g., \citealt{con02},
section 3.1). The references above show that other accurate
approximations have been developed, but they are complex.

The purpose of this paper is to present a family analytic orbit
approximations for the logarithmic and other spherically symmetric, or
axisymmetric potentials. These orbit functions are simply powers of
precessing ellipses, and so, very simple indeed. Nonetheless, for
small-to-moderate eccentricities, they are also surprisingly
accurate. For the logarithmic potential, the approximate orbits
librate around accurate numerical solutions, but do not diverge from
them over time. Equally surprising, these approximate solutions extend
continuously across all members of a family of potentials that range
from the Keplerian potential to potentials with slightly rising
rotation curves. Although they do not provide complete analytic
solutions, their simplicity and accuracy allow them to fill a large
part of the role of simple analytic solutions in providing a readily
understandable conceptual picture of the nature of orbits in these
potentials and the relations between them.

\section{Equations for Core Plus Power-Law Halo Potentials}
\subsection{Potentials}

In this section I present the basic equations and dimensionless
parameters that will be used in the following sections. All of the
potentials used in this work give accelerations described by the
equation,

\begin{eqnarray}
\nonumber
g(r) = \frac{-GM_*r}{(\epsilon^2 + r^2)^n},\\
\mbox{with}\  M(r) = \frac{r^3}{(\epsilon^2 + r^2)^n}
M_*,\\
\nonumber
\mbox{so}\  M_* = 2^n M(\epsilon) \epsilon^{2n-3},
\end{eqnarray}

\noindent where $\epsilon$ is a core scale length and $M_*$ is the
constant scale mass of the potential. The last equation relates the
scale mass to the mass contained within the scale length. The
acceleration of equation (1) has many useful limiting forms. First of
all, when $ r \gg \epsilon$, the acceleration increases linearly with
radius, as does the circular orbit velocity. Thus, we have a
solid-body rotation at small radii. In the special case of $n = 3/2$
we have the well-known Plummer model. The potential corresponding to
this acceleration is also related to those studied in general
scattering theory (see e.g., \citealt{new66}).

Secondly, when $ r \gg \epsilon$, we have the power-law limit, $g
\propto 1/r^{n-1}$. It is convenient to define $n = 1 + \delta$, so
that the exponent of $r$ is $\delta$ in this limit. Three interesting
special cases are: 1. $n = 0$, $\delta = -1$, circular velocity
$v_{cir} \propto r$, 2. $n = 1$, $\delta = 0$, circular velocity
$v_{cir} =$ constant, and 3. $n = 3/2$, $\delta = 1/2$, circular
velocity $v_{cir} \propto r^{-1/2}$. Clearly, these cases correspond
to linearly rising, flat and Keplerian rotation curves, respectively.

The potential and density equations corresponding to the acceleration
of equation (1) are given by,

\begin{eqnarray}
\nonumber
\Phi(r) = \frac{-GM_*}{2\delta
\left(\epsilon^2 + r^2 \right)^\delta},\\
\rho = \frac{3M_*}{4\pi}\ 
\left[ \frac{\epsilon^2 + \left(\frac{1-2\delta}{3}\right) r^2}
{\left( \epsilon^2 + r^2 \right)^{2+\delta}} \right].
\end{eqnarray}

\noindent The latter equation shows that density is approximately
constant within the core. This equation yields negative densities when
$\delta > 1/2$, and $r^2 > (3/(2\delta - 1))\epsilon^2$, i.e., at
large radii in potentials where the outer falloff is more rapid than
the Keplerian one. Such potentials are not very relevant to galaxy
dynamics. However, some examples of them are used below, so it is
worth noting that the density is positive out to quite large radii for
potentials where $\delta$ is only slightly larger than $1/2$.

Currently, both observations (e.g., \citealt{dia05}) and numerical
structure formation models support the idea that halo profiles are
universal. Among the most popular forms are the \citet[henceforth
NFW]{nav97} and \citet{her90} profiles. The density functions of both
these profiles have a charateristic scale length, but they are also
singular at the center, so their scale length does not correspond to a
core size as in equation (2). Power-law potentials that are singular
at the center will also be considered below. However, there is
evidence that the core density profiles of some galaxies are flat
rather than singular (e.g., \citealt{don04} and references
therein). \citet{hen05} has recently argued that the NFW state is a
partially relaxed state, that in isolation ultimately relaxes to a
core profile with solid body rotation. (For other recent views on the
origin of halo profiles see the recent works of \citealt{an05},
\citealt{deh05}, and \citealt{lu05}.)

At large radii the NFW and Hernquist density profiles fall off as
$1/r^3$. This is also true of the profile of equation (2) when $\delta
= 1/2$. I would emphasize, however, that the goal of this work is not to study
how well a certain class of orbital approximations applies to specific
halo potentials, Rather it is to describe how this family of
approximations works in a range of simple schematic potentials, such
as those of equation (2).

\subsection{Dimensionless Variables}

It will be convenient to work in dimensionless variables. The
dimensionless radius is $x = r/\epsilon$, and the dimensionless time
is $\tau = t/{\tau_{ff}}$. ${\tau_{ff}}$ is a characteristic dynamical
time, i.e., the free fall time at $r = \epsilon$. In terms of these
variables, the Newtonian equation of motion can be written,

\begin{equation}
\frac{d^2x}{d\tau^2} = \frac{-{\mu}x}{(1 + x^2)^n},
\end{equation}

\noindent with the parameters,

\begin{equation}
\mu = \frac{GM_*\tau_{ff}}{\epsilon^{2n}},\ \  h = 
\frac{v_{\phi}r\tau_{ff}^2}{\epsilon^2},
\end{equation}

\noindent where $\phi$ is the azimuthal coordinate, $v_{\phi}$ is the
azimuthal velocity, and $h$ is the dimensionless specific angular
momentum.

To further simplify the equation of motion we use the usual Kepler
case substitutions of the inverse radius, $u = 1/x$, for the radius
and the azimuthal advance, $d\phi = (d\phi/dt) dt$ for the time
change. The resulting equation of motion is,

\begin{equation}
\frac{d^2u}{d{\phi}^2} + u - 
c \left( \frac{u^2}{1+u^2} \right)^{\delta - 1}
\left( \frac{u}{(1 + u^2)^2} \right) = 0,
\ \ c = \frac{GM_*}{h^2}.
\end{equation}

\noindent Henceforth, we will abbreviate $du/d\phi$ as $u'$. All of
the results below apply to limiting or special cases of this equation.

Before proceeding to these results we should briefly discuss the
magnitude of the constant $c$, which is the only parameter in equation
(5). Using the definitions of the parameters $\mu$ and $h$ (eq. 4) we
have,

\begin{equation}
c = \left( \frac{GM_*}{\epsilon^{2n-4}} \right)\ 
\frac{1}{(rv_{\phi})^2}.
\end{equation}

\noindent In the case of circular orbits this reduces to,

\begin{equation}
c = \frac{(1 + x_{cir}^2)^n}{x_{cir}^4},
\end{equation}

\noindent
where $x_{cir}$ is the dimensionless orbit radius. We see that for
$x_{cir} << 1, c \propto 1/ x_{cir}^4$, while for $x_{cir} >> 1, c
\propto 1/ x_{cir}^{4-2n}$, which decreases to zero at large radii for
potentials with $n < 1$. At $x_{cir} = 1, c = 2^n$, and thus, for orbits
of this size $c$ is of order a few in most of the potentials of
interest.

\section{The Logarithmic Potential}
\subsection{p-ellipse Solution to First Order in the Eccentricity}

We begin with the case of the logarithmic potential. The equation of
motion of the softened logarithmic potential is given by equation (5),
with $\delta = 0$. In the case of large orbits, or negligible
core softening length, the equation of motion reduces to

\begin{equation}
uu'' = c - u^2.
\end{equation}

\noindent
Henceforth, in this and all sections except 3.3, where the softening
is treated explicitly, we will take $\epsilon = 1$, $x = r$, and $u =
1/r$, as is the convention for power-law potentials.

Orbits derived from numerically integrating this equation look like
precessing ellipses. Like elliptical orbits they are bounded by two
turning-points (see e.g., \citealt{bin87}, \citealt{val05}). These
facts motivate the trial of a solution of the form,

\begin{equation}
u = \frac{1}{p} f(\phi) \left[ 1 +
e \cos \left( (1-b){\phi} \right) \right],
\end{equation}

\noindent
where $f$ is an unknown function used to track the deviations from the
ellipse which the remaining factors describe. That is, $e$ is the
eccentricity, the $b$ factor determines the precession rate, and the
semi-major axis of the ellipse $a$ is given by $p = a(1-e^2)$. $p$ is
the semilatus rectum (see e.g., Ch. 2 of \citealt{mur99}). (Valluri et
al.'s summary of Clairaut's rotating ellipse and other historical
approaches to the lunar theory are also relevant.)

The solution of equation (9) solves equation (8) up to terms of order
$e^2$, if we take $f$ equal to the inverse square root of the
ellipse. Specifically, this approximate solution is,

\begin{equation}
u = \frac{1}{p} \left[ 1 +
e \cos \left( (1-b){\phi} \right) \right]^{1/2},
\ \ \mbox{with,}\   
c_1 = \frac{1}{p^2},\ 
\mbox{and,}
\ b_1 = 1 - \sqrt{2}.
\end{equation}

\noindent
(For details see Appendix A.)

The second equality above gives the first order approximation $c_1$
for the constant $c$ of equation (8) in terms of $p$, or the semi-major
axis $a$. However, $p$ is now the ``semilatus rectum'' for an oval
that equals the square root of an ellipse. We will call a curve like
equation (10), that is a power of an ellipse, a p-ellipse, for
(precessing) power ellipse. Note that the parameter e functions in
some ways like the ellipse eccentricity, but not in all ways, see
discussions following equations (11) and (23) and in the
appendices. For convenience we will continue to call it the
eccentricity (see \citealt{ada05} for a discussion of eccentricity
definition). 

Many of the geometric properties of ellipses are retained by
p-ellipses. For example, the major axis of a p-ellipse from equation
(10) is given by a simple function of $p$ and $e$,

\begin{equation}
2a = r(\phi = 0) + r((1-b)\phi = \pi)
= p \left[ \frac{\sqrt{1+e} + \sqrt{1 - e}}
{\sqrt{1-e^2}} \right].
\end{equation}

\noindent
Then the combination $c_1 a^{2}$ is a simple function of $e$ whose
value only varies from 1.0-1.27 as $e$ is increased from 0 to 0.7.

The two elliptical focii are still important for p-ellipses. On the
other hand, the relation between the true anomaly and eccentric
anomaly angles, as conventionally defined in celestial mechanics (see
\citealt{mur99}), is much more complicated for p-ellipses than for
ellipses.

As conic sections, ellipses are tilted planar cuts of circular
cones. Similarly, p-ellipses can apparently be derived as tilted
planar cuts of circular paraboloids given by the formula $z^2 = (x^2 +
y^2)^{1/2 + \delta}$, though this is difficult to prove in general. In
the case of the logarithmic potential, the resulting Cartesian
equation for the $\delta = 0$ p-ellipse is a fourth order
polynomial. The Cartesian equation for the p-ellipses is not the same
as that for super ellipses or the Lam\'{e} curves (defined by the
equation $(x/a)^n + (y/b)^n = c$), nor the elliptic or hyper-elliptic
curves (which are quadratic in the variable $y$). These families of
curves have quite different shapes.

\begin{figure}
\epsscale{.8}
\plotone{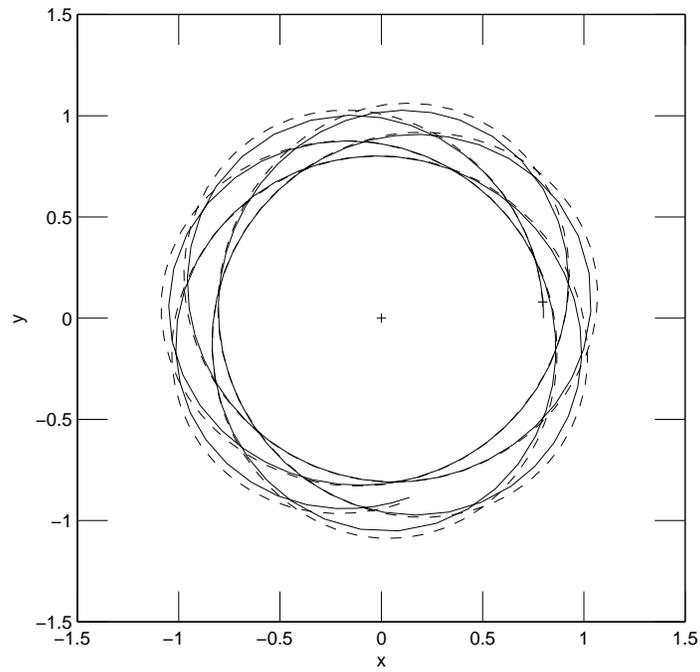}
\caption{Sample x-y orbit in the logarithmic potential with
energy/angular momentum constant $c = 1.2$ and eccentricity $e =
0.3$. The solid line is a numerical integration of the orbit, dashed
line is a first-order p-ellipse approximation. See text for
details. Note the end points of both in the lower right
quadrant.\label{fig1}}
\end{figure}

\begin{figure}
\epsscale{.8}
\plotone{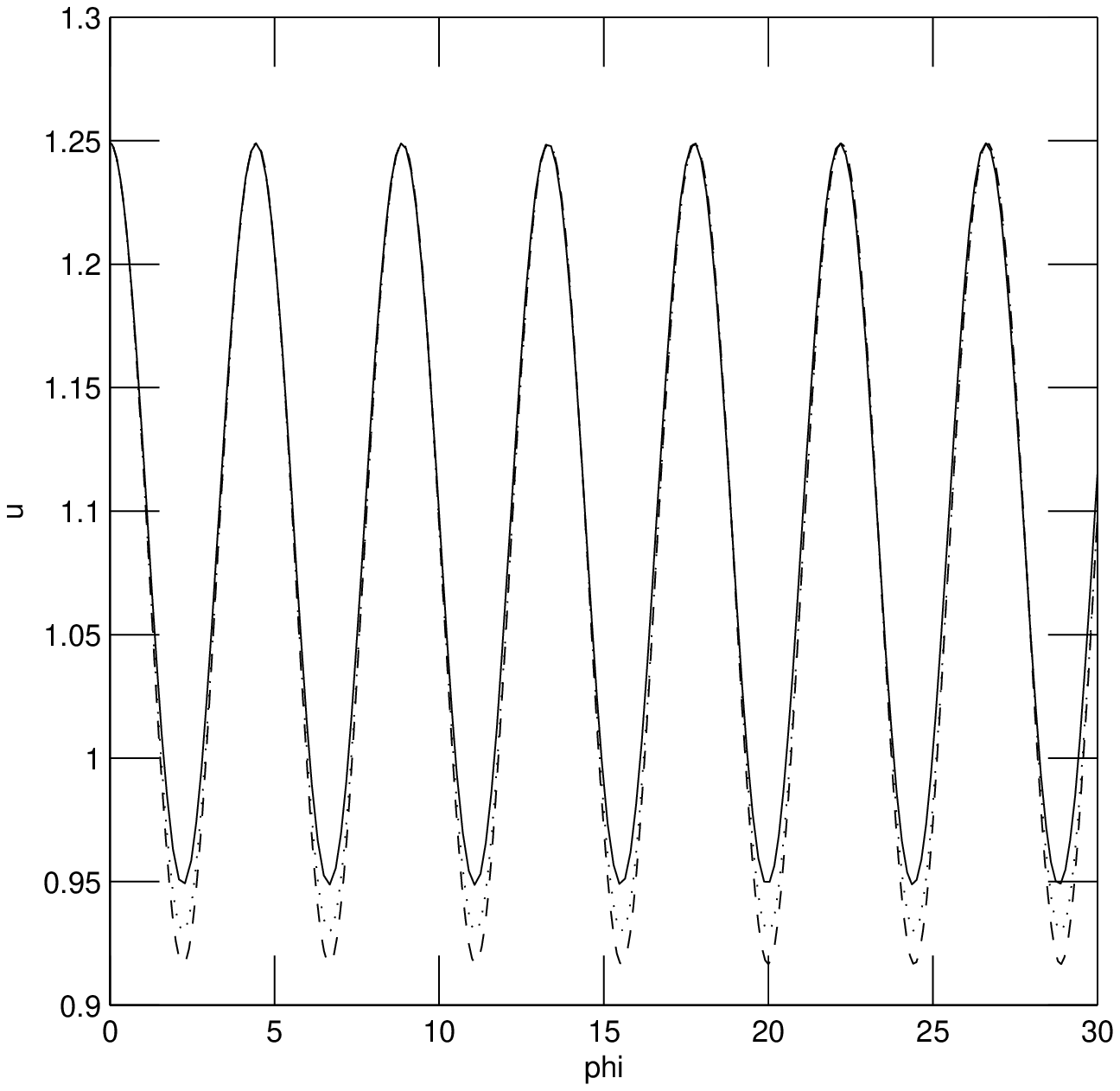}
\caption{Inverse radius variable u versus azimuthal angle $\phi$ for a
sample orbit in the logarithmic potential with energy/angular momentum
constant $c = 1.2$ and eccentricity $e = 0.3$. The solid line is a
numerical integration of the orbit, dashed line is the first-order
p-ellipse approximation, and dotted line is the second-order
approximation.\label{fig2}}
\end{figure}

The approximate solution given by equation (10) is technically only
good to first order in $e$, so we might expect it to have quite limited
accuracy, except for very small values of $e$. However, Figures
\ref{fig1} and \ref{fig2} compare the fit of equation (9) to a full
numerical solution of equation (8) in the case of a moderate
eccentricity of $e = 0.3$. Figure \ref{fig1} shows that the orbital fit
is excellent over several orbital periods. Figure \ref{fig2} shows
that the approximation does not systematically drift away from the
numerical solution after many cycles, and in fact is very close except
near periapse. This shows that the approximate value of the precession
factor $b$ is surprisingly accurate. (Part of the reason for this
accuracy is the slow variation of the constant $c$ with eccentricity
discussed above. These points will be discussed further below.)

\subsection{Second Order Approximation (p-ellipse plus epicycle)}

A clue to the lack of drift between the approximation (eq. (10)) and
the solution noted in the last subsection is that the only second
order terms that appear are proportional to $(c e^2)
\sin^2((1-b)\phi)$ and $(c e^2) \cos^2((1-b)\phi)$. Moreover, there
are no third or higher order terms. The fact that these are simple
sine or cosine squared terms suggests that we could cancel them with
more trigonometric functions in the approximate solution. Thus, we
next consider approximations consisting of the sum of a p-ellipse and
an elliptical epicycle (constant plus cosine term),

\begin{equation}
u = \frac{1}{p_1} \left[ 1 +
e \cos \left( (1-b){\phi} \right) \right]^{1/2}
+ \frac{1}{p_2} \left[ 1 +
e_2 \cos \left( (1-b){\phi} \right) \right].
\end{equation}

As before, we substitute this into equation (8) and equate the sum of
terms with common powers of $e \cos((1-b)\phi)$ to zero. Additionally,
we assume that $e_2 \propto e$. We find that the resulting constant
terms yield an expression for $c$, the first order terms yield the same
value of $b$ as in the previous subsection, and the second order terms
yield a relation between $e_2/e$ and $p_2/p_1$. If we further demand
that equation (12) equals equation (10) at $\phi = 0$ (and that both
equal the initial value $u(0)$ of the full solution), then we obtain
another relation between these two ratios and can solve for both. The
results are,

\begin{eqnarray}
\nonumber
c_2 = \left(\frac{1}{p_1} + \frac{1}{p_2}
\right) ^2 \left[ 1 - \frac{e^2}
{2p_1 \left( \frac{1}{p_1} + \frac{1}{p_2} \right)} 
\right],
\ b_2 = 1 - \sqrt{2},\ 
e_2 = \sqrt{1+e} - 1,\ 
\mbox{and,}\\
\frac{a_2}{a_1} =
\left( \frac{\frac{3}{4} - \frac{e_2}{e}}
{2 \left( \frac{e_2}{e} \right)^2} \right)  
\left[ \left( \frac{3}{4} - \frac{e_2}{e} \right) ^2
+ 2 \left( \frac{e_2}{e} \right)^2 \right] ^{1/2}.
\end{eqnarray}

\noindent The subscript ``2'' is used on the constants $b$ and $c$ to
indicate that these values are accurate to order $e^2$.

It is assumed that the added cosine term has the same precession
factor as the p-ellipse, so it does not drift from the p-ellipse
approximation. A fortuitous cancellation leads to a zero second order
correction to the precession factor $b$. We will see below that this is
not generally true in other potentials.

This second order solution is plotted along with the numerical and
first order solutions in the example of Figure \ref{fig2}, and we see
that it doesn't provide much improvement. The reason for this is that
in going from the first to the second approximation, we have changed
the higher order terms remaining in the orbit equation from two second
order terms that partially cancel at most azimuths, to more than half
a dozen third order terms, which do not generally cancel each
other. Note that in this example: $p_1 = 1.93, p_2 = 1.73$, and $e_2 =
0.14$.

These third order terms have coefficients that are third order powers
of sines and cosines, so they could presumably be eliminated by adding
still more trigonometric terms to the approximate solution. We will
not pursue the details, but it thus appears that we could improve the
approximation with the addition of more epicycles or Fourier
terms. Because the p-ellipse part of the solution is itself a good
approximation, and because successive Fourier coefficients contain
successively increasing powers of the p-eccentricity, the solution
series should converge steadily for moderate eccentricities.

\subsection{The Softened Logarithmic Potential}

The accuracy of the simple p-ellipse approximate solution for orbits
in the logarithmic potential is remarkable. In this and the following
subsections we will see that it is not unique. On the contrary, it is
useful over a range of interesting potentials. In this subsection we
consider cases of equation (5) with $\delta = 0$, but with nonzero
core length $\epsilon$ (see eq. (1)), corresponding to the
softened logarithmic potential.

In this case equation (5) can be written as,

\begin{equation}
uu'' + u^2 - c = -u^2 (uu'' + u^2),
\end{equation}

\noindent where the left-hand-side is the same as equation (8), and
the additional nonlinear terms have been placed on the
right-hand-side. (Note that in this subsection we return temporarily
to the convention of measuring all length units relative to
$\epsilon$, i.e, here $u = {\epsilon}/r$.) The term in parenthesis on
the right-hand-side of equation (14) is also very similar to equation
(8), which is a large part of the reason that the p-ellipse
approximations continue to work well in this case.

Upon substitution for $u$ from equation (10), and equating zeroth and
first order terms in $e\cos((1-b)\phi)$ as above (see Appendix A), we
derive new forms for the factors $b$ and $c$ appropriate to this
potential,

\begin{equation}
c_1 = \frac{1}{p^2} 
\left( 1 + \frac{1}{p^2} \right) ,
\ \ b_1 = 1 - \left[ 2 \left(
\frac{1 + \frac{2}{p^2}}{1 + \frac{1}{p^2}}
\right) \right]^{1/2}.
\end{equation}

\noindent
In the limit of orbit sizes much larger than the softening length ($p
\gg 1$), these expressions reduce to those of equation (10), as they
must. Additionally, in this case, we have two other limiting
cases. The first is the limit of small orbit sizes where the rotation
curve is linearly rising. In that case, $b_1 \simeq -1$, which as we
will shortly see is very interesting. The other special case is the
exchange or rotation curve downturn region, defined by $p \simeq 1$,
where there are also interesting resonances.

Before proceeding to discuss these cases it should pointed out that if
we substitute the second order approximation of equation (12) into
equation (14), we find that as in the previous subsection, there is no
second order correction to the $b$ factor. The $c$ factor, however,
does have a correction of order $e^2$,

\begin{eqnarray}
\nonumber
c_2 = \left( \frac{1}{p_1} + \frac{1}{p_2}
\right)^2  \left( 1 + \left(
\frac{1}{p_1} + \frac{1}{p_2} \right)^2 \right)
\left[ 1 + \frac{e^2(1-b)^2}
{4 p_1 \left( \frac{1}{p_1} + \frac{1}{p_2} 
\right)} \right],\\
\ \ b_2 = 1 - \left[ 2 \left(
\frac{1 + 2 \left( \frac{1}{p_1} + \frac{1}{p_2}
\right)^2}
{1 + \left( \frac{1}{p_1} + \frac{1}{p_2}
\right)^2}
\right) \right]^{1/2}.
\end{eqnarray}

\begin{figure}
\epsscale{.9}
\plotone{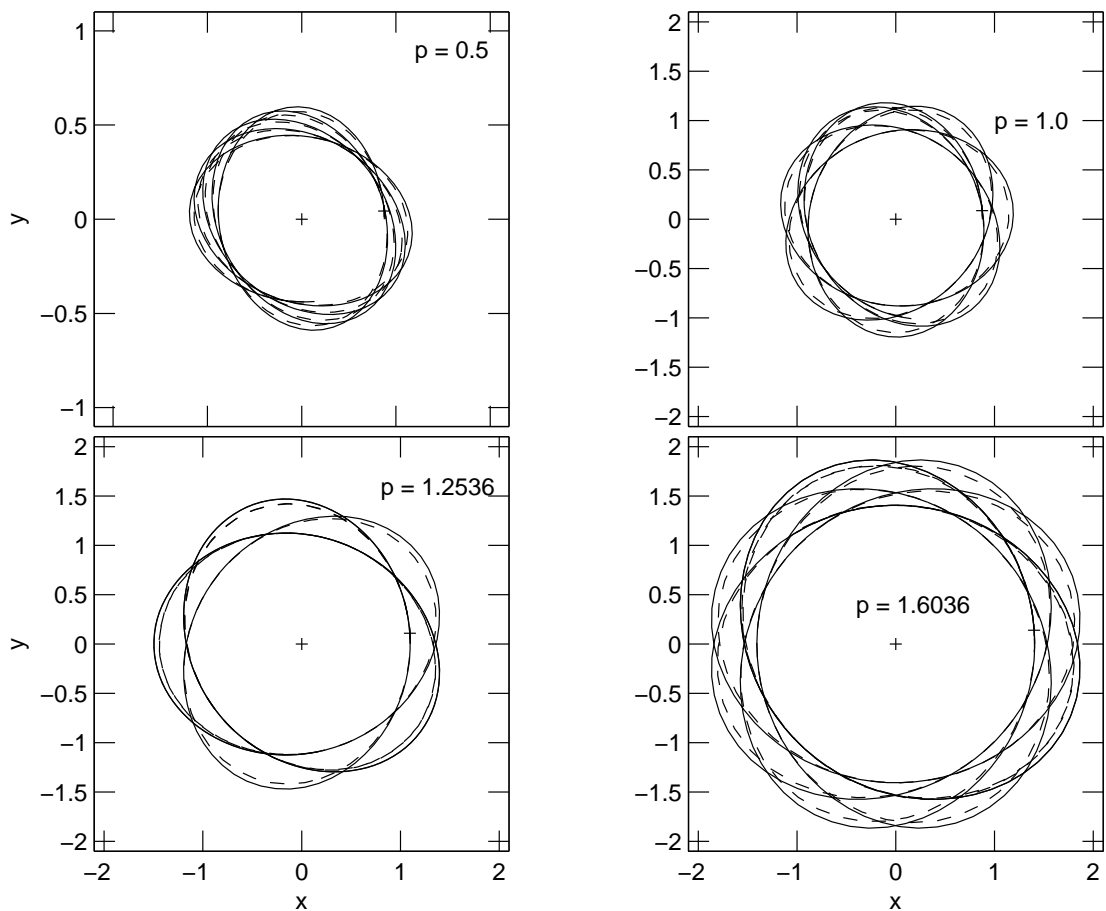}
\caption{Four sample x-y orbits in the softened logarithmic potential
with orbit sizes (semi-latus rectum) of $p = 0.5, 1.0, 1.2536$ ($b =
-2/3$), and $1.603567$ ($b = -3/5$). The p-ellipse eccentricity $e =
0.3$ in all cases. The solid line is a numerical integration of the
orbit, dashed line is the first-order p-ellipse approximation. Note
scale difference in the upper left panel relative to the others. See
text for details.\label{fig3}}
\end{figure}

Figure \ref{fig3} illustrates the goodness of fit of the (first order)
p-ellipse solution, for different size orbits. As we would expect from
the relation $b \simeq -1$, panel a) shows that the analytic solution
is a slowly precessing, nearly closed, oval when $p$ is small (here $p
= 1/2$). The panel also shows that the analytic approximation fits the
numerical solution quite well. It librates around it, but returns
periodically, just as in the case of the pure logarithmic potential.

The second panel of Figure \ref{fig3} is much like Figure \ref{fig1},
which corresponds to the large $p$ limit in the present case. That is,
the solution is a more rapidly precessing, open p-ellipse, which the
analytic approximation tracks quite well. It is interesting that we
can see this behavior already at $p = 1$, the case shown.

The third and fourth panels of Figure \ref{fig3} show special cases of
resonant orbits in the transition region. The third panel shows the
case $b = -2/3$, corresponding to $p \simeq 1.2536$. Both analytic and
numerical curves close (or nearly close) after 3 times around the
center. The fourth panel shows the case $b = - 3/5$, corresponding to
$p \simeq 1.6036$, and we have more loops before closure.

The factor $b$ ranges over the interval $(-1, 1 - \sqrt{2})$. The
infinite rational numbers in this interval correspond to closed
(analytic) orbits. However, if we write a rational number m/n, there
are very few in this interval with small values of m or n, i.e., low
order period resonances. Like the above examples most of these will
have radii of $p \simeq 1$. We reiterate that the value of $b$ does
not depend on the eccentricity up to at least second order. The
examples of Figure \ref{fig3} have a moderate eccentricity of $e =
0.3$.

In sum, Figure \ref{fig3} shows that the first order p-ellipse
approximation provides a good fit to orbits in this potential, as in
the case of the logarithmic potential. In the previous subsection we
found that the second order approximation yielded only a small
improvement. In the present case the second order derivation involves
a very unwieldy, high order polynomial in $p_1$ and $p_2$. Since this
analysis produces no significant new results, I will not present it
here (except for eq. (16)).

\section{Potentials with Rising or Falling Rotation Curves}

\subsection{General p-ellipses}

In the preceding section we considered the special cases of bound
orbits in the logarithmic or softened logarithmic potentials, i.e.,
where $\delta = 0$ in equation (5). In this section we consider
positive and negative values of $\delta$, which correspond to falling
and rising rotation curves. We will not, however, consider the most
general form of equation (5). To simplify the analysis and the
discussion, we will neglect softening and consider only pure power-law
potentials. (We will also return to conventional variables, where $u =
1/r$.)

In this case the equation of motion becomes,

\begin{equation}
uu'' = cu^{2\delta} - u^2.
\end{equation}

\noindent The simple p-ellipse solution in this case is,

\begin{equation}
u = \frac{1}{p} \left[ 1 +
e \cos \left( (1-b){\phi} \right) 
\right]^{\frac{1}{2} + \delta},
\end{equation}

\noindent
with the variables defined as above. Specifically, we see that when
$\delta = 0$ we have the approximation of the previous section for the
logarithmic potential, and when $\delta = 1/2$ we have the ellipse for
the Keplerian case. Given that in both of these cases equation (18) is
an excellent approximation, it is not completely surprising that the
equation is also an excellent approximation for all intermediate
values of $\delta$. We will return to that point in a moment.

First we follow the procedure of the previous section. We substitute
equation (18) into equation (17) and eliminate all terms of zeroth and
first order in $e$, by making the identifications,

\begin{equation}
\ c_1 = \left( \frac{1}{p} \right)^{2(1-\delta)},
\ \mbox{and,}\ 
\ b_1 = 1 - \sqrt{2(1-\delta)}.
\end{equation}

\noindent
Comparison to equation (10) shows great similarity, with the addition
of some simple factors of $1 - \delta$. The quantities $p$, $e$, and
$\delta$ determine a p-ellipse, but it is also desireable to specify
an orbit in terms of the conserved specific angular momentum and
specific energy. Relations between these variables are presented in
Appendix B.

In Figure \ref{fig4} we show some sample approximate and numerically
integrated orbits with a range of $\delta$ values, and with $p = 1$,
$e = 0.2$ in all cases. A quick glance at this figure reveals a couple
of basic points. First, the structure of comparable orbits varies a
great deal over these different potentials. Second, the first order
p-ellipse approximation is very good in all cases. If we look more
closely at the individual panels we see that, as in the case of the
logarithmic potential, the approximation librates a short distance
from the numerical solution, only to return to it shortly
thereafter. As we will discuss in the next section, this is generally
true because the precession frequency implied by the second equality
of (19) is exactly right for all the potentials (i.e., the same as in
Newton's theorem).

We see that the orbit shown in Figure \ref{fig4}b is similar to that
of Figure \ref{fig1}. The exponent $\delta$ in this case is half way
between that of the logarithmic and Keplerian potentials, so the fact
that the orbit is similar to the former, but more slowly precessing,
reflects continuity between these solutions. Figure \ref{fig4}c is
very close to the Keplerian case, and the precession is very slow
indeed. If we cross to the other side of the Kepler case, as in Figure
\ref{fig4}d with $\delta = 0.7$, both approximate and numerical
solutions appear more erratic. This example is in the range between
the Kepler solution and the Cotes spiral solution at $\delta = 1.0$ in
the present notation. (The MATLAB routine ODE45 was used throughout
this paper.)  In any case, we don't expect the p-ellipse approximation
to be very useful for values of $\delta$ of about or greater than 1.0.

Figure \ref{fig4}a shows an example of going past the logarithmic
potential to those with negative values of $\delta$. The exponent $1/2
+ \delta$ of equation (18) becomes very small in the vicinity of
$\delta = -1/2$, so even with significant values of the eccentricity
$e$, the radial excursion of orbits is very small. Nonetheless, the
figure shows that the p-ellipse approximation continues to be good.

The two panels of Figure \ref{fig5} show sample orbits at still more
negative values of $\delta$. The first panel shows a cases where
$\delta = -1.1$ is very close to that of the solid-body potential, and
so is quite similar to Figure \ref{fig3}a. The second panel shows a
still more extreme case ($\delta = -1.5$), and here the deviations
between approximate and numerical orbits become quite
significant. Thus, it seems that the p-ellipse approximation is at
best only qualitatively useful for values of $\delta$ much more
negative than the solid-body case.

The normalized (by a factor of $p^2$) second order terms that remain
when the p-ellipse is substituted into the equation of motion provide
an estimate of the inaccuracy of that approximate
solution. Technically, they provide an estimate of the inaccuracy of
the $u''/u$ value, but given the common periodicity of $u$ and its
derivatives, this is roughly equivalent to the relative inaccuracy in
$u$. All the second order terms are proportional to
$(e\cos((1-b)\phi))^2$. If we take the average value of 1/2 for the
cosine term, a first rough estimate of the error is $e^2/2$, which,
when compared to the numerical results, turns out to be about right in
the vicinity of $\delta = 0$.

The $\delta$ dependence of this error estimate is given by the
function: $(1+2\delta)(1-\delta)(\delta^3 + 0.5\delta^2 + 3.5\delta -
1)$. When $\delta$ values vary from -0.5 to 0.5, the absolute values
of the function are always less than 1.2, and mostly in the range
0.5-1.2, i.e., of order unity. For values $\delta \ll -0.5$, the
absolute value of the function rises rapidly (eventually as
$(1+2\delta)\delta^4)$, explaining the diminished accuracy of the
p-ellipse approximation. Within the domain of $\delta$ values of the
most interest (-0.5 to 0.5), eccentricity is the dominant factor in
the error estimate. Section 5.2 offers a way to derive more accurate
p-ellipse approximations at large eccentricity.

The approximate nature of the first order p-ellipses, and these
limitations on their domain of accurate application, distinguish them
from a universal analytic solution. However, the latter evidently
doesn't exist, and the p-ellipses provide accurate orbit
approximations in power-law potentials over most of the domain of
interest for studies of galaxy dynamics.

\subsection{Second Order Approximation for Power-Law Potentials}

As in the case of the logarithmic potential, we can attempt to
eliminate terms of second order in $e$ by adding the $p_2, e_2$ terms
of equation (12) to equation (18). In the more general cases of
$\delta \ne 0.0$ there are more second order terms than in the
logarithmic potential, so we might expect the correction to be more
useful than in that case.

However, direct comparison of the second order approximation to
numerical integrations in cases like those shown in Figures \ref{fig4}
and \ref{fig5} shows that the situation is not that simple. I find
that for $\delta$ values in the range $0 < \delta < 1/2$ there is
little difference between the p-ellipse and second order
approximations, as in the case of the logarithmic potential. For $-1/2
< \delta < 0$, the second order approximation fits the numerical
orbits significantly better than the first order p-ellipse. For
$\delta < -1/2$, the amplitude of the second order approximation fits
better, but its precession rate is apparently not accurate because the
orbital phase drifts quite rapidly. We conclude that, as in the case
of the logarithmic potential, there is not much reason to prefer the
second order approximation in the case of general power-law
potentials.

The following equations define the second order solution,

\begin{eqnarray}
\nonumber
\nonumber
b_2 = 1 - 
\left[ \frac{2(1-\delta)}
{1 + \frac{\delta}{2} \left( 1 - 4{\delta}^2 \right) 
\left( \frac{p_2}{p_1 + p_2} \right) e^2}
\right] ^{1/2},
\\
\nonumber
e_2 = \left( 1+e \right)^{\frac{1}{2}+\delta}
- 1, \ \ \ (initial\  condition),
\\
\nonumber
\frac{p_1}{p_2} = \left( \frac{(1 + 2\delta)
\left( \frac{3}{4} - \frac{e_2}{e} \right)}
{2 \left( \frac{e_2}{e} \right)^2} \right)
\left[ 1 \pm (1 + 2\delta) \left[
\left( \frac{3}{4} - \frac{e_2}{e} \right)^2
+ 2  (1 - 2\delta) \right]^{\frac{1}{2}} \right],
\\
p_1 = \left( 1 + \frac{p_1}{p_2} \right)
c^{\frac{-1}{2(1-\delta)}},
\end{eqnarray}

\noindent
where the variables $e, c,$ and $\delta$ are given, and the above
equations are solved for $p_1, p_2, e_2$, and $b_2$. The equation
labelled $initial\ condition$ is derived by equating the initial value
of the first and second order approximations. 

The expression for $b_2$ in equation (20) has a slight dependence on
$e$. This allows the second order approximation to drift relative to
the first order solution. It also allows for the possibility of a
rational value of the precession period relative to the orbital
period, and closed orbits, at special values of $e$. We will consider
this further in Sec. 5.

\subsection{Potentials with Closed Orbits}

The second equality of (19) allows for values of $\delta$ that yield
rational values of $b_1$, and thus, closed orbits for all
low-to-moderate values of $e$ in those potentials, in the first order
p-ellipse approximation. These potentials are generally isolated,
i.e., not dense in the mathematical sense, on the $\delta$ axis,
though there are exceptions. The existence of these closed orbits and
their basic structure can also be deduced from the apsidal precession
theorem and its generalizations. However, there is the question of how
small the eccentricity must be for Newton's theorem to be valid, and
the generalizations of \citet{val05} to high eccentricities are
complex. If p-ellipse formulae like those of eqs. (19) and (20) are
sufficiently accurate, then they provide simple alternatives for
guidance on these questions. We examine this question in this
subsection.

From the second equality of (19) we deduce that in order for $b_1$ to
have a rational value, we must have,

\begin{equation}
\delta = 1 - \frac{1}{2} \frac{j^2}{k^2},
\end{equation}

\noindent
where $j$ and $k$ are integers. Then, $b_1 = (k-j)/k$. As a simple and
generalizable example consider the sequence $k$ = 1, 2, 3,..., with $j
= k + 1$. The corresponding sequence of $b_1$ values is $b_1 =
-1/k$. Table 1 gives $\delta$ values for some elements of this
sequence.

Figure \ref{fig6} gives examples of orbits in this sequence with
eccentricities of $e = 0.2$ and orbit sizes of $p =
1.0$. Specifically, the $k$ = 2, 3, 4, and 20 orbits are shown. The
first panel shows a very simple closed orbit, that with $\delta =
-1/8$ is close to the logarithmic potential. The analytic and
numerical solutions are very similar to each other. To keep the plot
readable we only show a couple of times around the orbit. However, a
plot like that of Figure \ref{fig2} for many orbits shows no drift
between the numerical and analytic curves. I will not attempt to prove
that the orbits in this potential are closed, but it appears that for
practical purposes they are. (This has been checked at other values of
$e$ as well.)

The second panel of Figure \ref{fig6} corresponds to a slightly
falling instead of a rising rotation curve with $\delta = 1/9$. It is
slightly more complex, but still closed, and with excellent agreement
between the analytic and numeric curves. As before, there is no drift
between the two curves after many orbits.

The trends of more complexity in the closed curves, and good agreement
between analytic and numeric curves, continue in the third panel of
Figure \ref{fig6}, where $\delta = 7/32$.

The fourth panel, where $\delta \simeq 0.45$, shows that things get
very interesting as we approach the Kepler case. The orbits become
very slowly precessing p-ellipses that eventually close. As k advances
to large values in this sequence, $b_1$ goes to zero, and according to
equation (21) $\delta$ goes to 1/2. Thus, the sequence converges on
the Kepler case. In contrast to the sparsity of potentials in this
sequence in the region of the logarithmic potential, the convergence
of this sequence shows that such potentials do densely populate the
$\delta$ axis at values just less than the Kepler value. These
statements are also true for a second sequence defined by $k$ = 1, 2,
3,..., with $j = k + 2$, and so on.

Thus, the properties of the p-ellipses, originally derived as simple
analytic approximations, have revealed families of potentials with (at
least practically) closed orbits, like the Kepler potential, and
unlike the logarithmic potential. Appropriately, these families
cluster thickly on one side of the Kepler potential. Interestingly,
these special potentials include the $\delta = -1/8$ and $1/9$
potentials, but not the simple $\delta = 1/8$ and $1/4$ potentials. This
is explained by the precession law of the p-ellipses.

I have avoided discussing the very simple case of $k = 1$, $j = 2$,
and $b_1 = \delta = -1$. This is the first element of the sample
sequence above, and is the solid-body potential. The analytic and
numeric curves for this highly resonant case do not agree. This is
true in spite the fact that we have shown good agreement in examples
above, which have $\delta$ values a little less than -1.0.

\section{Discussion and Applications}

\subsection{Apsidal Precession and Integrals of the Motion}

It has been noted already that one of the secrets of the success of
the first-order p-ellipse approximations is that their precession
rates accurately match those of the true orbits in power-law
potentials up to moderate eccentricity values. It remains to
demonstrate explicitly that they satisfy Newton's apsidal precession
theorem, and consider large eccentricities.

As to the former, we note that the minimum of u (apoapse) occurs when
$(1-b)\phi = \pi$, the maximum (periapse) when $(1-b)\phi = 0$, and
the difference is,

\begin{equation}
\Delta \phi = \frac{\pi}{\sqrt{2(1-\delta)}},
\end{equation}

\noindent
which is the apsidal precession theorem for p-ellipses. \citet{val05}
remind us that with an acceleration proportional to $1/r^{N-3}$,
Newton's apsidal precession theorem gives $\Delta\phi =
\pi/\sqrt{N}$. When we compare this form of the acceleration to the
one used in this paper (with no softening), we find $N = 2(1-\delta)$,
so the two expressions for the apsidal precession are the same. This
proves the that the first-order p-ellipses satisfy the apsidal
precession theorem. (Note that with eq. (15) we also can readily
derive the apsidal precession as a function of orbit size in the
softened logarithmic potential.)

\citet{val05} emphasized that apsidal precession depends on orbital
eccentricity at large eccentricities, so the Newton formula and
equation (22) are not correct in that limit. These authors derived
integral expressions for the dependence of the precession rate on N
and $e$ for the power law potentials. The first equation of (20)
provides a simple approximate formula for this dependence.

Before comparing this simple formula to the Valluri et al. results, we
should recall that the p-ellipse eccentricity we have used up to this
point is not the same as the usual definition of the eccentricity
(except in the case of the inverse square law). If an orbit is assumed
to be approximately a precessing ellipse, then one definition of
eccentricity is in terms of the periapse distance using the formula
$r_{min} = a (1-e)$. We will call the eccentricity defined in this way
the classical eccentricity, and denote it $e_o$. The classical
eccentricity and the p-ellipse eccentricity used above are related by
the equation,

\begin{equation}
\left( 1-e \right)^{\delta + \frac{1}{2}} 
= 1 - e_o.
\end{equation}

\noindent
The classical eccentricity is generally significantly less than the
p-ellipse eccentricity, though they are nearly the same for potentials
close to the Kepler potential.

With that notational disparity accounted for, we return to the topic
of the dependence of precession on eccentricity, and the comparison of
Valluri et al.'s results to the $b_2$ expression in equation
(20). First of all, it should be emphasized that we have seen that the
second-order approximation described above was not found to be highly
accurate in many cases. Valluri et al.'s semi-analytic apsidal
precession angles, illustrated in their Figure 2, are more accurate
than those derived from equation (20).

Nonetheless, the $u-\phi$ plot in Figure \ref{fig7} shows an example
where the precession of a numerical orbit solution after 10-20 orbits
is much better fit by the second order precession rate than the first
order rate. (Their libration amplitudes are comparable.) Equation (20)
does seem to be an improvement over the $b_1$ formula of equation
(19). The example shown in Figure \ref{fig7} has $e = 0.95$ and $e_o =
0.84$, so it is quite eccentric. The example in Figure \ref{fig7} has
$\delta = 1/9$, a closed orbit value, but at this large eccentricity
the orbits are not closed (see Fig. \ref{fig8}).

The $b_2$ expression of (20) also captures and helps us to understand
many of the qualitative features of Valluri et al.'s results. First of
all, the dependence on eccentricity is quite weak in equation (20),
and so, it only becomes important for large values of $e$ (unless
$\delta$ is very large). Indeed, either equation (20) or Valluri et
al.'s graphs show how good an approximation the constant precession
rate $b_1$ is for given values of $e$ and $\delta$ (and thus overcome
the primary objection to the use of the apsidal precession theorem up
to modest eccentricities). Secondly, both formulations predict that
the precession rate is constant at all eccentricities for the
logarithmic potential. Thirdly, they both agree that the apsidal angle
$\Delta\phi$ increases (decreases) with eccentricity when $\delta < 0$
($\delta > 0$). In fact, equation (20) provides a fair quantitative
approximation to the plots of Valluri et al. at values of $\delta$ not
far from zero.

Valluri et al. also state that their work gave no evidence for
``isolated cases of zero precession as e was increased.'' The $b_2$
expression in (20) indicates that such cases might exist at values of
$\delta$ just slightly less than 1/2, but even if this prediction is
confirmed, these cases would be very anomalous. Another somewhat
arcane prediction of equation (20) and Valluri et al.'s results is
that potentials very near the resonant ones discussed in the previous
section should have a closed orbit at one specific, high value of the
eccentricity. These potentials have smaller values of $\delta$ than
the resonant one for positive $\delta$.

The discussion of this subsection clearly relates to the classical
integrals of the motion. The precession rate $b$ (or the precession
frequency divided by the orbital frequency), is an integral of the
motion, much like the initial orbital phase angle which is often used
as such (see Sec. 3.1 of \citealt{bin87}). By indicating the presence
of closed orbits, the expressions for $b$ in (19) and (20) tell us
(approximately) when these integrals are isolating.

\subsection{An Alternate Approach to p-ellipse Orbit Approximation}
\subsubsection{Better Approximations at Larger Eccentricities}

The p-ellipse orbit approximations above were derived from an analytic
perturbation analysis, with some implicit constraints. In this section
we learn more about these approximations by relaxing some of these
constraints. Figure \ref{fig7} shows how, in the case of an orbit of
high eccentricity ($\delta = 1/9$ and $e = 0.95$), the p-ellipse
approximations drift in period relative to the numerical
solution. (This is not surprising given the \citet{val05} results on
apsidal precession.)  This motivates an attempt to improve the
p-ellipse solution by changing the value of $b$ to match the numerical
result. In the case shown in Figure \ref{fig7} this means changing
from the analytic value of $b_2 = -0.3027$ to -0.3558.

According to equation (19) or (20), a different $b$ factor corresponds
to a different $\delta$ value. In this case (using eq. (19)), we go
from $\delta = 1/9 = 0.1111...$ to $\delta^{\prime} \simeq
0.0809$. Another difficiency of the p-ellipse approximation shown in
Figure \ref{fig7} is the radial range. With the revised value of
$\delta$ we can also revise $p$ and $e$ to fit the numerical maximum
and minimum radius. The revised p-ellipse approximation is given by
the equation,

\begin{equation}
\frac{1}{r} = 1.1105\left[ 1 + 0.7026 \cos(1.3558{\phi}) 
\right]^{0.5809}.
\end{equation}

\noindent
This provides a very good fit to the orbit as shown in Figure
\ref{fig8}, where the numerical orbit is shown by the solid line, and
the approximation by the dashed line.

To recapitulate, we have treated the p-ellipse quantities as free
parameters, first adjusting the precession rate to match the numerical
value, and then the orbital size and eccentricity to match the radial
range. The $\delta$ value was also changed to maintain the
relationship of equation (19). This procedure seems to yield an highly
accurate orbit approximation for such loop type orbits, even at quite
high eccentricity.

\subsubsection{p-ellipse Transformation and Perturbation Approximation}

The result of the previous subsection can be viewed as achieving a
more accurate fit to a high eccentricity orbit than the first-order
p-ellipse by using a smaller eccentricity p-ellipse orbit taken from a
different (power-law) potential. This suggests that there is an
approximate mapping between p-ellipses, representing an approximate
symmetry of the power-law potentials. The nature of this relation is
that the orbits in a given potential at high eccentricity correspond
to orbits of lower eccentricity in a different potential, with the
same precession rate. Of course, the first-order p-ellipse
approximation fits the lower eccentricity orbit better.

As an aside, note that this relation suggests that periodic orbits may
be found at high eccentricity in a potential with generally aperiodic
orbits, when the correspondence is to an individual orbit in a
periodic potential.

However, the assumption that the first order equation (19) holds at
high eccentricity is likely to be a rather crude ansatz. Another
approach is to return to the perturbation analysis of Sec. 4.1, but
now allow the $\delta$ factor of the p-ellipse approximate solution to
be different than that of the equation of motion, call it
$\tilde{\delta}$. We will still fix the $b$ factor to the numerical
result, because even a small precessional drift will result in a large
difference between the approximate and true orbit.

Then, requiring the cancellation of zeroth and first order terms in
$e\cos((1-b)\phi)$ yields the following relation,

\begin{equation}
(1-b)^2 = \frac{2(1-\delta)}
{\left[ 1 - \frac{\delta}{2} (1-4\tilde{\delta}^2) e^2
\right]},
\end{equation}

\noindent
(Note the similarity to the expression for $b_2$ in eq. (20))

For a given potential exponent $\delta$ and precession parameter $b$,
this gives a relation between the exponent $\tilde{\delta}$ of the
corresponding potential, and the eccentricity of the p-ellipse in that
potential. Requiring that the approxmation also fit the maximum and
minimum radii of the numerical orbit (which can be derived from the
conserved energy and angular momentum, see Appendix B) gives three
relations for the variables $\tilde{\delta}$, $e$, and $p$. Second
order terms in the perturbation expansion do not cancel, but since the
$b$ value is correct, this procedure is better than the direct
second-order approximation. The results for the case shown in Figure
\ref{fig8} are: $\tilde{\delta} = -0.003$, $p = 0.87$, and $e = 0.77$.
The values are very similar to those of equation (24). This
perturbation approach is less ad hoc than that used to derive equation
(24), and thus, provides a more solid basis for approximating high
eccentricity orbits.

\subsubsection{Epicycloid Comparison}

A specific comparison to the epicycloid approximation of \citet[see
their eqs. (20) and (21)]{ada05} is also shown in Figure \ref{fig8},
as a dotted curve. This curve was constructed by requiring that it
have the same maximum and minimum radii, and the same precession rate
as the numerical solution. It can be seen from Figure \ref{fig8}, and
the corresponding $u - \phi$ graph that it deviates from the numerical
solution a bit more at intermediate radii than the specially fitted
p-ellipse.

Overall, the simple versions of both approximations seem to be about
comparably accurate in the case shown. It seems likely that the
epicycloids, like the p-ellipses, can provide good orbit
approximations in the potentials studied above. \citet{ada05} have
considered the properties of the epicycloid orbits in general
potentials, so I will not examine this issue in detail in this
paper. The advantages of the p-ellipses include their simple form in
polar coordinates, and the fact that their orbit parameters change in
very simple ways as functions of the conserved quantities and the
parameter $\delta$ in power-law potentials.

\subsection{Galaxy Collisions, Bars and Precession Period Resonance}

In this paper, I will not present any detailed applications of the
p-ellipse theory developed above. Indeed, since the orbits have long
been studied numerically in many relevant cases we do not expect new
applications, as much as new perspectives on old
applications. Nonetheless, in this section I will highlight a couple
of applications that seem worthy of further development.

The first is simply the application of p-ellipses to obtain the
general characteristics of satellite orbits in potentials with either
rising or falling rotation curves. The simplicity of p-ellipse allows
one to quickly produce a wide range of possible satellite orbits in
any particular situation. In studies of interacting galaxies, one can
then use the impulse approximation to estimate the effect of a close
encounter on initially circular stellar orbits in the disk of the
primary. The initial effects of such encounters are kinematic, with
self-gravitational effects taking time to accumulate, so solutions of
the perturbed orbit equations would provide a reasonable picture of
the early development of tidal structures (see review of
\citealt{str99}). This, in turn, would be a very efficient means of
constraining the satellite orbit and the primary gravitational
potential needed to produce observed tidal features, at least in
encounters with modest dynamical friction.

The second is based on the slowly precessing orbits in the inner
regions of the softened logarithmic potential discussed in Sec. 3.3
and illustrated in Figure \ref{fig3}a. Equation (15) tells us how the
precession $b$ depends on orbit size $p$ in this case. In particular,
for values of $p < 1$, we have $b_1 \simeq -1$, which is the slow
precession. If a disturbance (e.g., tidal) elongates initially
circular orbits in this region, the slow precession tells us that they
will stay approximately aligned for some time. This in itself is not
remarkable in a potential that is approximately solid-body in that
region. However, the p-ellipse equations not only tell us about this
kinematic bar, but they also tell us quantitatively how this structure
would precess into a spiral at somewhat larger radii. Of course they
do not include the effects of self-gravity on these kinematic waves.

A third application concerns resonant orbit-satellite
interactions. \citet{tre84}, for example, have
emphasized the importance of resonant orbits in galaxy bar dynamics
and in dynamical friction effects on satellites. Frequently, when
resonances are discussed in galaxy dynamics they are Lindblad
resonances. These are local (in radius), and usually visualized
in the epicyclic approximation as small integer ratios between the
epicyclic period and the period of the mean (guiding center) orbit.

The existence of closed orbits and nearly closed orbits described
above naturally raises the topic of resonant interactions in galaxy
disks involving these orbits. One example would be resonances between
eccentric p-ellipse type orbits and circular orbits with resonant
periods, and radii close to the apsides of the eccentric
orbit. Because of the radial range of the eccentric orbit, such
interactions would not be as localized as the Lindblad resonances. For
non-Keplerian potentials, Kepler's third law generally has a
(complicated) dependence on eccentricity (see Appendix C), which will
make the cataloguing of such resonances somewhat difficult. These
resonances seem qualitatively similar to those between orbit families
in barred potentials, so like the latter they may play a role in
building bulges, especially at early times.

Period resonances involving the precession of p-ellipse orbits may
also be important. Figures \ref{fig6}a,b remind us that potentials
close to the logarithmic potential with such orbits have precession
periods of about a few times the orbital period. In these nearly flat
rotation curve potentials, this means that the precession periods of
disk stars are about equal to the orbital periods of satellites with
orbital radii a few times larger. Higher order resonances would occur
at larger satellite orbital radii. Thus, especially in the early
stages of galaxy evolution, this interaction could play a very
important role in the evolution of disks and satellite orbits. It may
also be important in galaxy interactions. The immediate objection is
that the closed orbit potentials described above are very sparsely
distributed on the $\delta$ axis, i.e., very rare. A counter-argument
is that potentials with orbits that are nearly closed, i.e., do not
drift much in phase over a duration of several precession periods,
span finite intervals on the $\delta$ axis near the closed orbit
potentials.

Resonant orbit-precession interactions might also be important
between forming planets in accretion disks if the disk mass is great
enough to modify the Kepler potential, as seen by planets on
elliptical-like orbits. As described above, there are many closed
orbit potentials with $\delta$ slightly greater than the Kepler value
of 1/2. The precession rate is slow for these potentials, but the
evolutionary times extend over at least thousands of orbital periods,
so this is not a problem.

A similar application may be found for the orbits of stars in dense
clusters around massive black holes in galaxy nuclei. If the central
black hole dominates, but the distributed mass of the star cluster is
non-negligible, the effective potential will have a rotation curve
that is slightly flatter than a Keplerian one. This is again in the
region of dense closed orbit potentials, so precessional resonance
interactions with satellite star clusters or other massive black holes
may be important. As usual with resonant interactions, they likely
involve only a small minority of stars, and so, specially designed
N-body simulations are needed to study the phenomenon.

\section{Summary and Conclusions}

In the sections above, we have shown that curves derived as powers of
ellipse functions, called p-ellipses, provide simple, yet surprisingly
accurate approximations to orbits in a range of power-law potentials,
including the well-known logarithmic potential, and softened power-law
potentials. For a given power-law potential the p-ellipse function is
nearly as simple as an ellipse, but the family of p-ellipses extends
continuously across a physically interesting range of potentials.

There are at least two reasons why p-ellipses are good orbital
approximates across this range of potentials, and are likely to be the
best simple, analytic functions to do so. First, p-ellipses match the
tendency for orbits of a given energy and angular momentum to become
more circular as the potential changes from Keplerian to solid-body
(and $\delta$ decreases). In this sense the p-ellipses adjust well to
the appropriate orbital shape in a given potential. Secondly, the
precession rate obtained by demanding that the p-ellipse satisfy the
equation of motion in a given potential to first order in eccentricity
is the identical to that given by Newton's theorem. Thus, the
p-ellipses precess correctly. Moreover, a second order approximation
yields an eccentricity dependence of the precession rate that is in
qualitative agreement with the \citet{val05} semi-analytic results.

A number of the results obtained above for p-ellipses, like the
apsidal precession rates in power-law potentials, are not new, and
good approximations to individual orbits can be obtained with
epicycles or numerical integration. However, a family of simple curves
like the p-ellipses allow us to readily see trends across a range of
potentials, so they provide a simple, conceptual picture for orbital
variations, as discussed in the introduction.

Moreover, as shown in the previous sections the p-ellipses provide a
very powerful tool for studying characteristics like the occurrence of
(non-circular) closed orbits as a function of eccentricity in
different potentials. The key feature of the p-ellipses in this regard
is a relatively simple approximate formulae for precession rates as a
function of $\delta$ and $e$. Newton's theorem is valid in the limit
of small eccentricity, and Valluri et al.'s extensions involve
integrals that must be evaluated numerically. Equations (19) and (20),
though approximate, offer convenience. (Also see eq. (25).)

The example of the softened power-law potentials holds out the hope
that the p-ellipses can provide useful orbit approximations in other
non-power-law potentials. Non-axisymmetric potentials have not been
considered in this paper, but it seems reasonable to hope that
p-ellipses could provide good approximations to loop orbits in such
potentials. This issue deserves more study.

In Section 5.3 I described several examples of how the systematics of
p-ellipses might shed light on important astrophysical problems. This
conclusion will be even more general if p-ellipses prove to be good
orbital approximations in more types of potential.

Orbit theory is more general than celestial mechanics, and the
p-ellipse approximations should also be relevant to any field than
involves orbits in general potentials. Electron orbits in general,
steady (gradient) electric and magnetic fields are obvious examples.

In sum, there seems to be room for a great deal more development of
the theory and application of these simple curves. Their simplicity
may allow us to address a number of complex issues that would be hard
to study directly through the accumulation of numerical examples.



\acknowledgments

This research was partially supported by NASA Spitzer grant
1263961. This research had made use of NASA's Astrophysics Data System
abstract service. I am greatful for helpful input on an earlier version
from Bev Smith, Scott Tremaine, and S. R. Valluri. The referee, Fred
Adams, provided many suggestions that significantly improved the
paper. 






\appendix

\section{Some Perturbation Analysis Details}

The purpose of this brief appendix is to provide a few more details,
and a slight extension of the perturbation analysis leading to the
results of equation (10). This is the simplest of several such
calculations in the main text, but is representative of the common
procedure.

That procedure consists of substituting the adopted solution (first
equality of eq. (10)) into the equation of motion (8), and gathering
terms of common order in the factor $e\cos((1-b)\phi)$. Then the
requirement that these terms cancel up to a given order is used to
derive relations between the paramenters. In this case, the zeroth
order terms yield the condition,

\begin{equation}
c = \frac{1}{p^2} \left( 1 -
\frac{(1-b)^2 e^2}{4} \right),
\end{equation}

\noindent
for the constant $c$ of equation (8) in terms of the variables $b$,
$p$, and $e$. Note that since this is an expansion in terms of
$e\cos((1-b)\phi)$, the $e^2$ term (with no cosine factor) is included
in the zeroth or constant condition. This factor is not included in
the (first-order) results given in equation (10).

Terms of first order in $e\cos((1-b)\phi)$ yield the condition,
\begin{equation}
\frac{1}{2} (1-b)^2 - 2 + cp^2 = 0,
\end{equation}

\noindent
which, in combination with the previous condition, yield an expression
for the $b$ factor.

The terms of second order in $e\cos((1-b)\phi)$ are,

\begin{equation}
1 - \frac{(1-b)^2}{4} =
\frac{1}{2} - \frac{e^2}{4},
\end{equation}

\noindent
where a common factor of $1/p^2$ has been dropped, and the result for
$(1-b)$ from the previous condition has been used. With no extra
parameter to adjust, these terms do not cancel, but do provide an
estimate of the p-ellipse accuracy. It is interesting that while all
second order terms are small at small eccentricity, the term above is
smallest at large eccentricity. This is not the case for other
power-law potentials. 

\section{Relations Between Orbital Parameters}

In this appendix I present some relations between orbital parameters,
especially those between the semi-major axis $a$, the eccentricity
parameter $e$, and the specific angular momentum $h$ and specific
energy $\mathcal{E}$. The relation between $h$ and the constant $c$ is
given in equation (4), and equation (19) relates the latter to the
semi-latus rectum $p$, for a first-order p-ellipse. Combining these
yields, 

\begin{eqnarray}
\frac{h^2}{GM_*} = p^{2(1-\delta)}
= \left( \frac{a}{g_o(e)} \right) 
^{2(1-\delta)},\\
\mbox{with,}\  g_o(e) = 
\frac{1}{2} \left[ \frac{(1-e)^{1/2 + \delta}
+ (1+e)^{1/2 + \delta}}{(1-e^2)^{1/2 + \delta}} \right].
\end{eqnarray}

\noindent The second equality in (B1), $h$ versus $a$, is derived from
a generalization of equation (11) for the case $\delta \ne 0$.

The specific energy is defined as,

\begin{eqnarray}
\mathcal{E} = \frac{1}{2} \left( v^2 - 
\frac{GM_*}{\delta r^{2\delta}} \right),\  
\mbox{for}\ \delta \ne 0,\\
\mbox{and,}\ \mathcal{E} = \frac{1}{2} \left( v^2 -
GM_* \ln(r) \right),\ \mbox{for}\ \delta = 0,
\end{eqnarray}

\noindent where $v$ is the total relative velocity of satellite. The
radial velocity, $v_r$, can be derived in terms of the azimuthal
velocity by differentiating equation (18) with respect to time. Then,
the azimuthal velocity can be eliminated using the relation $h =
rv_\phi$. After some simplification, for the case with $\delta \ne 0$,
we get,

\begin{eqnarray}
\mathcal{E} = \frac{GM_*}{2a^{2\delta}}\ g_{\mathcal{E}}(e,\delta),\\
\mbox{with,}\ \ 
 g_{\mathcal{E}}(e,\delta) = g_o^{2\delta} (1+e)^{1+2\delta}
\left[ 1 - \frac{1}{\delta}(1+e)^{(\delta-1)(1+2\delta)} \right].
\end{eqnarray}

For the case $\delta = 1/2$, equations (A1), (A2), (A5), and (A6)
reduce to the usual Keplerian elliptical orbit relations: $\mathcal{E}
= -GM/2a$ and $h^2 = GMa(1-e^2)$.

For the case $0 < \delta \ll 1/2$ (i.e., nearly flat, but slightly
decreasing rotation curve), we have,

\begin{eqnarray}
g_o \simeq \frac{1}{1-e_o},\ 
g_{\mathcal{E}} \simeq 1 + e - \frac{1}{\delta}
\simeq -\frac{1}{\delta},\\
h^2 = GM_*a^2(1-e_o)^2,\ 
\mbox{and},\ 
\mathcal{E} = \frac{-GM_*}{2\delta a^{2\delta}},
\end{eqnarray}

\noindent where $e_o$ is the ``classical'' eccentricity of equation
(23). In contrast to the full relations, these limiting cases can be
easily inverted to derive $a, e$ as functions of $h, \mathcal{E}$.

\section{Time-Azimuth Relations and Kepler's Law for p-ellipses in
Power-Law Potentials} 

Relations between time and azimuth for p-ellipse orbits in power-law
potentials are not treated in the main text. Since these relations are
likely to be very important for any application involving p-ellipses
we derive them here. Specifically, we can derive an invertible
analytic expression for the logarithmic potential, and a implicit
relation for potentials with small values of $\delta$. To begin, the
equation for the conserved specific angular momentum gives the
azimuthal frequency, $d\phi/dt = h/r^2$. Next, we substitute the
p-ellipse solution for r to get,

\begin{equation}
\frac{d\phi}{dt} = \frac{h}{p^2}
\left[ 1 + e \cos \left( (1-b)\phi \right) \right]^{1+2\delta}.
\end{equation}

In the case of the logarithmic potential, $\delta = 0$, this can be
integrated to obtain,

\begin{equation}
\frac{h(1-b)t}{2p^2} \sqrt{1-e^2} =
\tan^{-1} \left\{ \sqrt{\frac{1-e}{1+e}} \left[ 
\tan \left( \frac{(1-b)\phi}{2} \right) -
\tan \left( \frac{(1-b)\phi_o}{2} \right)
\right] \right\},
\end{equation}

\noindent
where, on the right hand side, we assume that there is an initial
azimuth, $\phi_o$, on the left we assume that the initial time is $t =
0$. 

By setting $(1-b)\phi = 2\pi$ and $\phi_o = 0$ we get an expression
for the orbital period, $T$, in the logarithmic potential,

\begin{equation}
T = \frac{p^2}{h(1-b)}\ \frac{2\pi}{\sqrt{1-e^2}}.
\end{equation}

Using equations (10) for $b$, (11) for $p$ in terms of the semi-major
axis $a$, and (B1), (B2) for $h$ (with $\delta = 0$), we derive
Kepler's third law for first order p-ellipse orbits in the logarithmic
potential,  

\begin{equation}
T^2 = \frac{2\pi^2}{GM_*} 
\left[ \frac{2} {\left( \sqrt{1+e} + \sqrt{1-e} \right)^3}
\right]^2 a^2.
\end{equation}

\noindent
The eccentricity dependence of the expression in square brackets is
less than 10\% different than the $e = 0$ value when $e < 0.7$, and so
is quite weak.

For power-law potentials near the logarithmic, $|\delta| \ll 1$, the
procedure above gives the following integral equation, 

\begin{equation}
\frac{ht}{p^2} = \int^{\phi}_{\phi_o}
\frac{d\phi}{\left[ 1 + e\cos\left((1-b)\phi\right)
\right]^{1+2\delta}}
\simeq
\frac{1}{1-b} \int^{\phi}_{\phi_o}
\frac{\left[ 1 - 2\delta e \cos\left((1-b)\phi\right) \right]}
{\left[ 1 + e\cos\left((1-b)\phi\right)\right]}
d\left( (1-b)\phi \right),
\end{equation}

\noindent
where the last equality gives a first order approximation in
$\delta$. This equation can be further reduced to,

\begin{equation}
\frac{ht}{p^2} = 
\left( \frac{1-2\delta}{1-b} \right) I_o
- 2\delta(\phi - \phi_o), 
\ \mbox{with,}\ \
I_o = \int^{\phi}_{\phi_o}
\frac{d\phi}{\left[ 1 + e\cos\left((1-b)\phi\right)
\right]}
\end{equation}

\noindent
The integral $I_o$ is the same one that appears in the case of the
logarithmic potential, and whose solution is give in equation
(C2). The above equation cannot be solved explicitly for
$\phi$. However, an explicit form of Kepler's third law can be
derived; it is,

\begin{equation}
T^2 = \frac{2\pi^2}{GM_*}
\left[ \frac{1 - 2\delta - \delta\sqrt{1-e^2}}
{\sqrt{1-\delta}} \right]^2
\left[ \frac{2(1-e^2)^{\delta}}
{\left( (1+e)^{\frac{1}{2} + \delta} + (1-e)
^{\frac{1}{2} + \delta} \right)^3}
\right]^2 a^{2(1+\delta)}.
\end{equation}

\noindent
For small positive values of $\delta$, this expression gives an $e$
dependence that is not quite as weak as that of the logarithmic
potential. 




\clearpage



\begin{figure}
\epsscale{.9}
\plotone{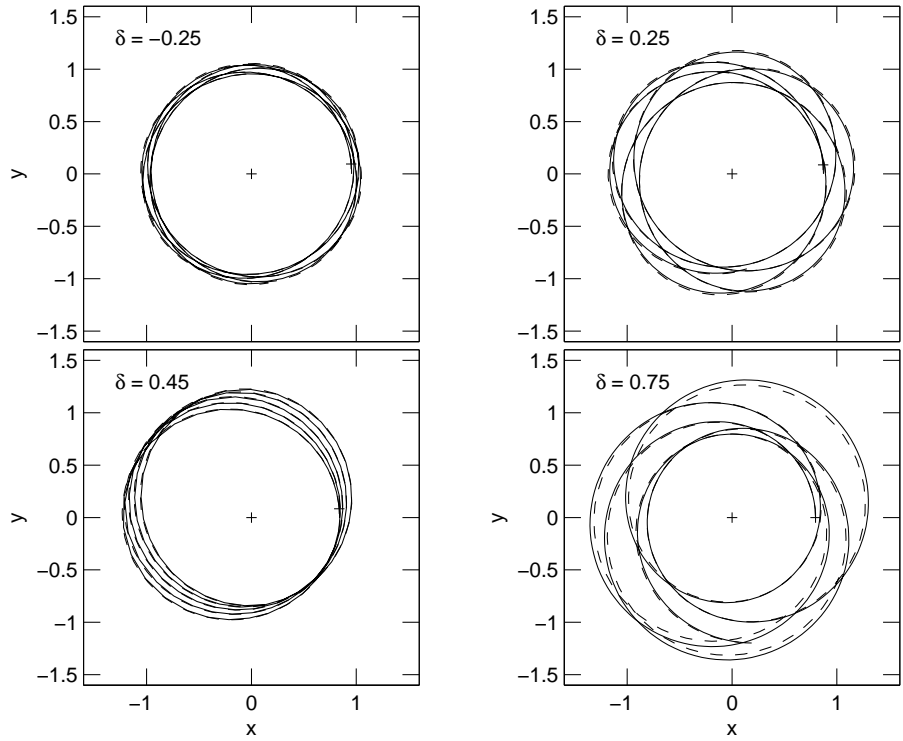}
\caption{Four sample orbits in several power-law potentials with
exponent $\delta$ values of $\delta = -0.25, 0.25, 0.45, 0.75$. The
p-ellipse eccentricity $e = 0.2$ and orbit size $p = 1$ in all
cases. The solid line is a numerical integration of the orbit, dashed
line is the first-order p-ellipse approximation.\label{fig4}}
\end{figure}

\begin{figure}
\epsscale{.9}
\plotone{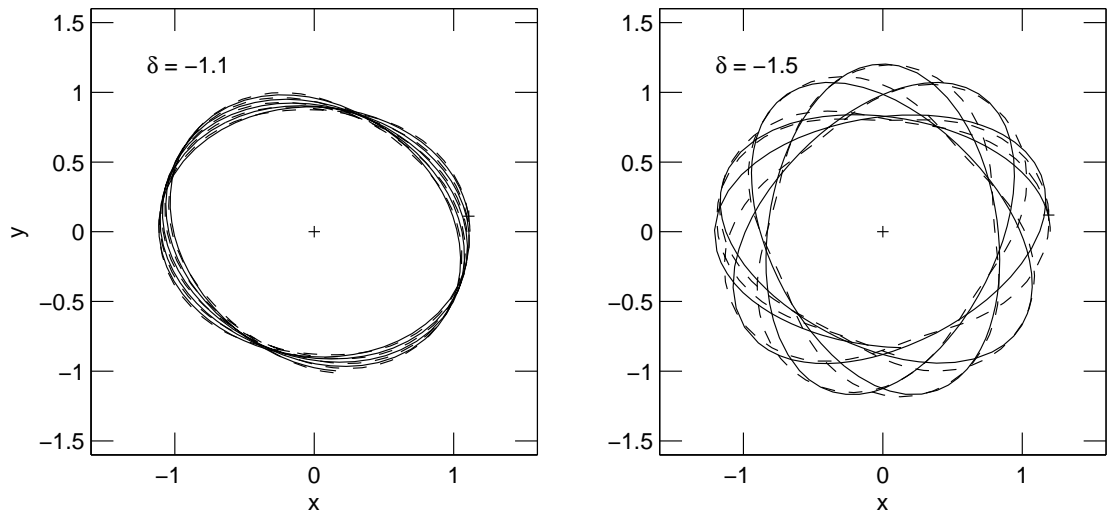}
\caption{Two sample orbits in power-law potentials with steeply rising
rotation curves, i.e., exponent $\delta$ values of $\delta = -1.1,
-1.5$. The p-ellipse eccentricity $e = 0.2$ and orbit size $p = 1$ as
in the previous figure. The solid line is a numerical integration of
the orbit, dashed line is the first-order p-ellipse
approximation.\label{fig5}}
\end{figure}

\begin{figure}
\epsscale{.9}
\plotone{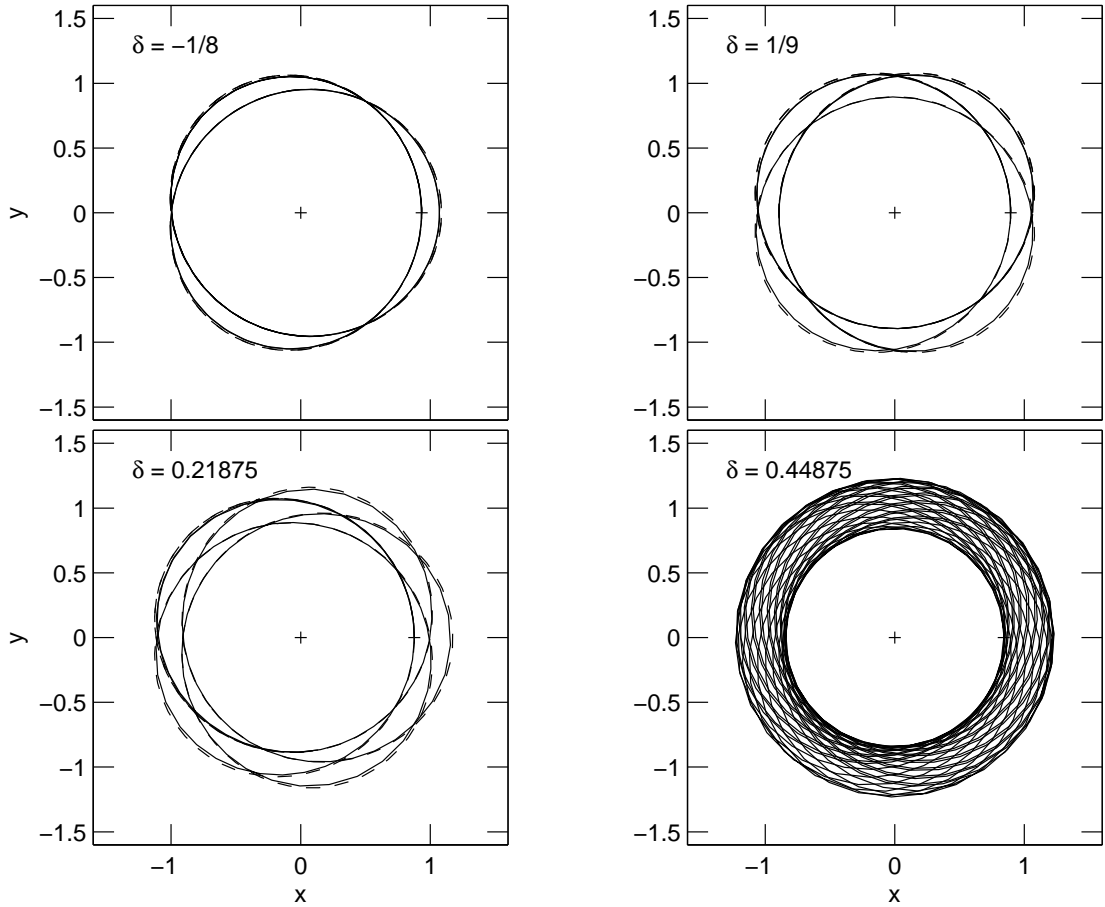}
\caption{Four sample closed orbits in power-law potentials with
exponent $\delta$ values of $\delta = -1/8, 1/9, 7/32, 0.44875$. The
p-ellipse eccentricity $e = 0.2$ and orbit size $p = 1$ in all
cases. The solid line is a numerical integration of the orbit, dashed
line is the first-order p-ellipse approximation. In the lower right
panel only the numerical curve is shown for clarity, though the
approximation agrees as well as in the other cases.\label{fig6}}
\end{figure}

\begin{figure}
\epsscale{.9}
\plotone{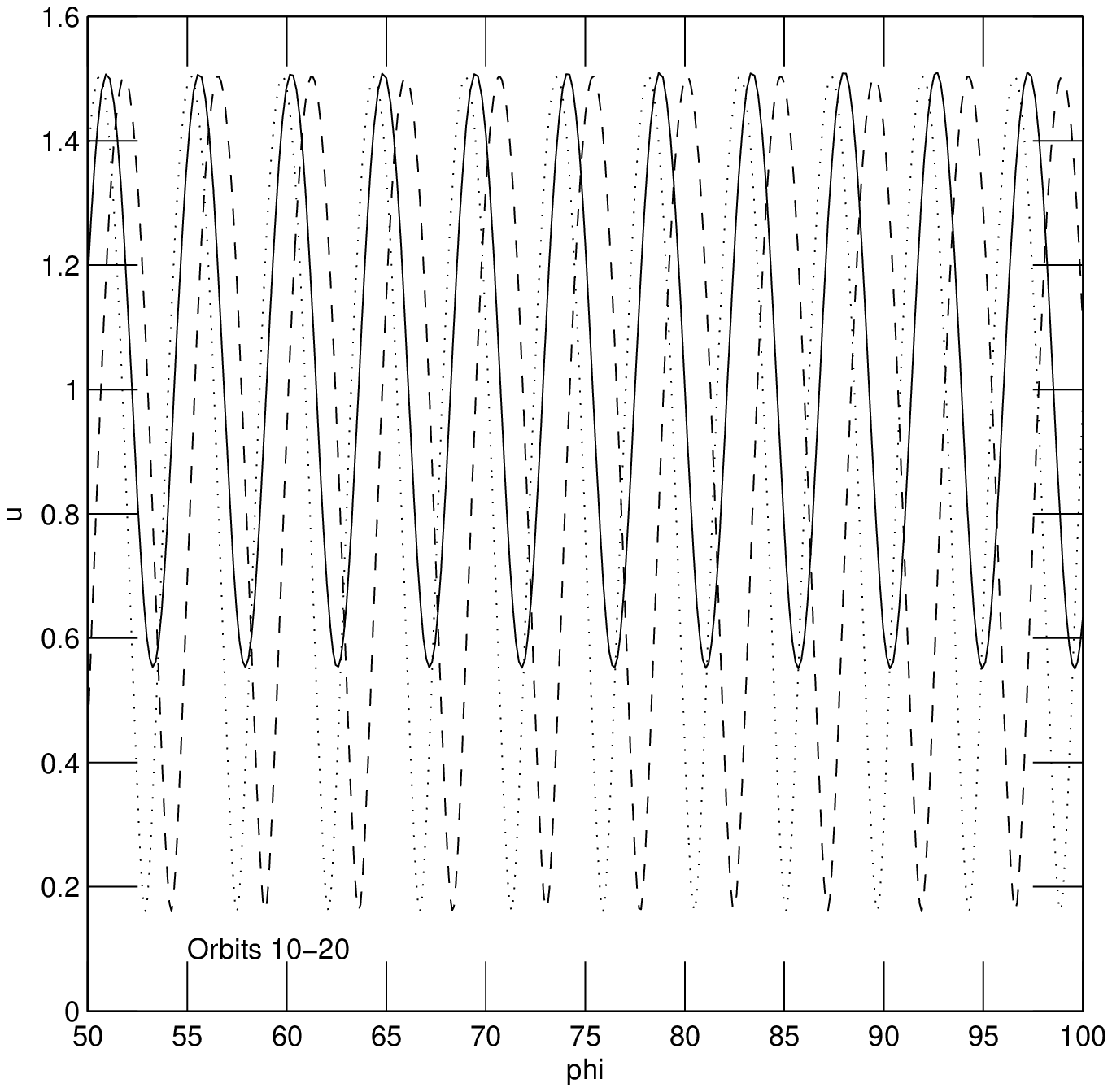}
\caption{Inverse radius variable u versus azimuthal angle $\phi$ for a
sample orbit in the $\delta = 1/9$ potential with energy/angular momentum
constant $c = 1.0$ and eccentricity $e = 0.95$. The solid line is a
numerical integration of the orbit, dashed line is the first-order
p-ellipse approximation, and dotted line is the second-order
approximation.\label{fig7}}
\end{figure}

\begin{figure}
\epsscale{.9}
\plotone{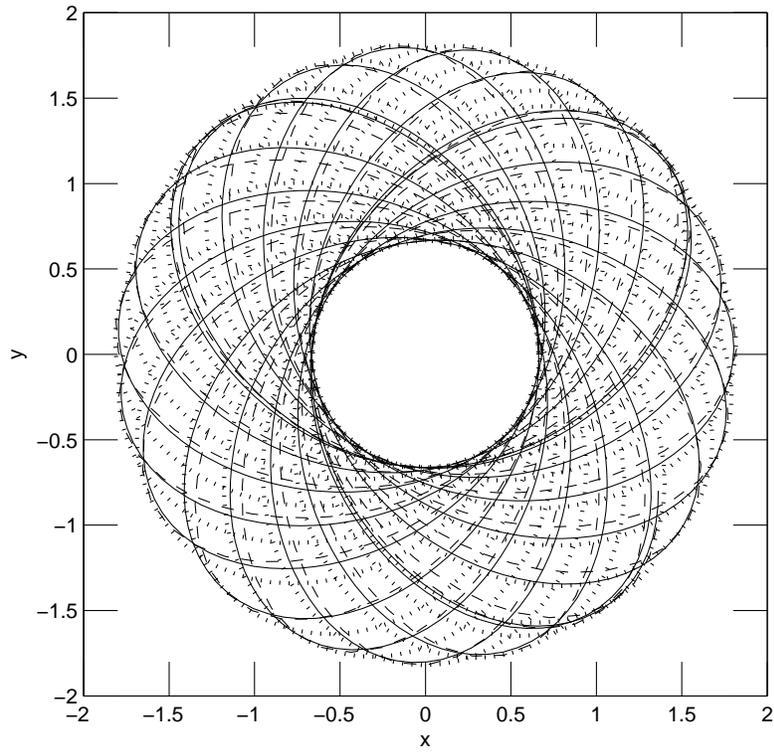}

\caption{Sample x-y orbit in the $\delta = 1/9$ potential with
energy/angular momentum constant $c = 1.0$ and eccentricity $e =
0.95$. The solid curve is a numerical integration of the orbit, dashed
curve is the first-order p-ellipse approximation, and dotted curve is a
spirographic approximation as described in the text.\label{fig8}}

\end{figure}

\clearpage

\begin{table}
\begin{center}
\caption{Examples of Potentials with Closed Orbits of Low
Eccentricity.\label{tbl-1}}
\begin{tabular}{crrrr}
\tableline\tableline
k & j = 1+k &  
\multicolumn{1}{c}{b = 1 - j/k\tablenotemark{a}} &
\multicolumn{1}{c}{$\delta$\tablenotemark{a}} \\
\tableline

1 &2 &-1 &-1\\
2 &3 &$-\frac{1}{2}$ &$-\frac{1}{8}$\\
3 &4 &$-\frac{1}{3}$ &$\frac{1}{9} = 0.11\overline{1}$\\
4 &5 &$-\frac{1}{4}$ &$\frac{7}{32} = 0.21875$\\
5 &6 &$-\frac{1}{5}$ &$\frac{7}{25} = 0.28$\\
6 &7 &$-\frac{1}{6}$ &$\frac{23}{72} = 0.3194\overline{4}$\\
7 &8 &$-\frac{1}{7}$ &$\frac{17}{49} = 0.346939$\\
8 &9 &$-\frac{1}{8}$ &$\frac{47}{128} = 0.367188$\\
... &... &... &...\\
20 &21  &$-\frac{1}{20}$ &$\frac{359}{800} = 0.44875$\\
... &... &... &...\\
& $p \simeq q$ &0 &$\frac{1}{2}$\\
\\

\tableline
k & j = 2+k & &\\
\tableline

1 &3 &-2 &$-\frac{7}{2}$\\
2 &4 &-1 &-1\\
3 &5 &$-\frac{2}{3}$ &$-\frac{7}{18} = -0.38\overline{8}$\\
4 &6 &$-\frac{1}{2}$ &$-\frac{1}{8}$\\

\tableline
\end{tabular}


\tablenotetext{a}{See text for definitions of b and $\delta$}
\end{center}
\end{table}

\end{document}